# Why They're Worried

## Examining Experts' Motivations for Signing the 'Pause Letter'

Isabella Struckman
&
Sofie Kupiec
Massachusetts Institute of Technology
istruck@mit.edu, skupiec@mit.edu

**Disclaimer:**

This paper was written for Alfred Spector's MIT Spring 2023 course 6.S963 Beyond Models – Applying Data Science/AI Effectively. It has not been peer reviewed, and its content does not necessarily reflect his or the authors' views.

Its goal is to present perspectives on the state of AI, as held by a sample of experts. These experts were early signatories of the Future of Life's recent open letter, which calls for a pause on advanced AI development. Utmost effort was put into accurately representing interviewees' perspectives, and they have all read and approved of their representation. However, no paper could offer a perfect portrayal of their position.

We feel confident in what opinions we do put forward, but we do not hold them tightly. In such dynamic times, we feel that no one should be resolved in their expectations for AI and its future.

# Table of Contents

# 1. Introduction

*"What sphinx of cement and aluminum bashed open their skulls and ate up their brains and imagination?*
*Moloch! Solitude! Filth! Ugliness! Ashcans and unobtainable dollars! Children screaming under the stairways! Boys sobbing in armies! Old men weeping in the parks!*
*Moloch! Moloch! Nightmare of Moloch! Moloch the loveless! Mental Moloch! Moloch the heavy judger of men!*
*Moloch the incomprehensible prison! Moloch the crossbone soulless jailhouse and Congress of sorrows! Moloch whose buildings are judgment! Moloch the vast stone of war! Moloch the stunned governments!*
- *Excerpt from 'Howl' (1956), by Allen Ginsberg*

Why do we do it? Why do we engage in deadly nuclear arms' races, play the tragedy of the commons on repeat, and, today, enter unflinchingly into an unslowable race to develop and publish increasingly complex and poorly understood artificial intelligence (AI) systems? In his 1956 poem 'Howl', Allen Ginsberg passionately introduces the contemptible beast Moloch, a monstrous physical embodiment of destructive societal pressures. Ginsberg says it is Moloch behind humanity's self-destructive tendencies. He pulls the strings when competitive and systemic pressures mix dangerously, and he is the reason we march down potentially destructive paths we never chose (Ginsberg, 1956).

On March 22nd, 2023, the Future of Life Institute (FLI) published an Open Letter calling for a six-month pause on "the training of AI systems more powerful than GPT-4". FLI asked the world to bask in a "long AI summer, [and] not rush unprepared into a fall." The short but highly publicized statement has, in fewer than two months, garnered over 50,000 signatures. The letter never mentions Moloch, but his presence echoes in the letter's description of AI today: "recent months have seen AI labs locked in an out-of-control race to develop and deploy ever more powerful digital minds that no one – not even their creators – can understand, predict, or reliably control" (Future of Life Institute, 2023).

A month after signing the letter, Max Tegmark–physicist, MIT professor, and president of FLI–directly cited Moloch in an interview with MIT colleague, Lex Fridman. Tegmark named him responsible for the current state of AI development. The breakneck pace at which generative models are currently published is directly caused by unavoidable competition driven by both industry and international affairs. He attributed this pressure cooker to none other than Ginsberg's dreaded Moloch (Tegmark & Fridman, 2023).

Both Tegmark and the letter-drafting institution he leads believe the current trajectory of AI development holds significant existential risk for humanity. However, this concern seems to not be the priority of some institutions that continue to develop and implement new state-of-the-art generative models at a rapid pace.

## 1.1. AI Today

AI has been ubiquitous in our daily lives (e.g., traffic navigation and facial recognition) for years. But the evolution of powerful generative AI—models that can generate novel content—has ignited what could be called a Moloch-driven 'AI arms race' between some of tech's biggest companies. Today, state-of-the-art models like Google's Bard (2023) and OpenAI's GPT-4 (2023) are easily able to both interact with and produce exceptionally realistic multimodal content. The public's overwhelming response to OpenAI's ChatGPT and DALL-E has forever changed industry. Thousands of companies are rapidly integrating generative AI into their operations, and competitors of OpenAI are developing competing models as fast as they can (Dastin, 2023). Benjamin Kuipers[1] described the mindset of the "major corporate actors that are building these large language models (LLMs). They are in an extremely intense competition with each other, and they are all very concerned with the fact that this could be a winner-take-all competition" (Kuipers & Kupiec, 2023).

As the intensity of this competition grows, so do concerns about the rapid adoption of these generative AI models. A subgroup of researchers and institutions, including Tegmark and FLI, think these risks may be substantial enough to threaten humankind.

## 1.2. The Letter

FLI is devoted to mitigating existential risks in AI which come "not from AI's potential malevolence or consciousness, but from its competence—in other words, not from how it feels, but what it does" (FLI, 2022). Institutions and individuals worried about existential risks fear that integrating powerful AI systems like GPT-4 into society could, through unexpected emergent behaviors and misaligned objectives, irreparably harm humankind. Though GPT-4 is technically "limited in its ability to form and execute complex plans, … it's hard to say that future releases won't have this ability" (Russell, 2023). FLI worries humans will be unable to maintain control over increasingly powerful AI systems, which may then do what they will (or, more specifically, what their objective function will encourage).

Consequently, FLI drafted a call for a pause in large model development so that the world's understanding of these systems may catch up. After five years of intense AI development, FLI claims it's time to equally focus on developing protocols to make AI "more accurate, safe, interpretable, transparent, robust, aligned, trustworthy, and loyal" (FLI, 2023). The open letter is very precise in its call. FLI does not seek to halt progress in AI or to affect most companies and researchers using and developing it. The letter highlights dangers specifically associated with large, opaque generative models, which are developed by only a few companies. The letter's thesis is helpfully bolded for its readers: "Powerful AI systems should be developed only once we are confident that their effects will be positive and their risks will be manageable" (FLI, 2023).

---

[1] Benjamin Kuipers is a Professor of Computer Science and Engineering at the University of Michigan. He was previously an endowed Professor in Computer Sciences at the University of Texas at Austin, where he served as Department Chair.

## 1.3. Public Reaction

The public reaction to this letter was explosive, and just about every argument for and against both its technical and political validity was made in the coming weeks (Kapoor & Narayanan, 2023). A flourishing mainstream discussion ignited over whether the letter's call for a pause was a positive idea or even possible, whether it was focused on the right issues, and over the range of risks AI holds today and in the future. Responses, both mainstream and by experts, have spanned a wide spectrum: from intense disagreement to enthusiastic support.

The breadth of responses to the letter speaks to its substantial public impact. Many responses were thoughtful, poignant, and exceptional contributions to essential global conversations on how (and if) AI can be carefully introduced to the world. Considerable nuanced attention was paid to the letter, the institution that wrote the letter, and even the identities of its signatories.

However, almost no attention has been paid to most of the signatories. The opinions of only a few public-facing signatories, like Elon Musk and Max Tegmark, have entered the public consciousness. FLI reached out personally to and received signatures from hundreds of top researchers and industry participants in the field weeks before the letter was published. But these signatories, faceless to many, have largely been placed into a generic "letter supporter" bucket. Expert signatures gave credibility that drove the letter's rise to fame, but it's unclear whether the signatories are truly as aligned with the letter as it is easy to assume.

## 1.4. Motivation

Do all 50,000 signatories of the letter believe with certainty in the existential risks of AI? Almost certainly not. Do the first few hundred? Those top experts, researchers, and technologists that FLI personally asked to sign the statement? Did these unconsulted signatories agree with every line as they read the contentious letter? The answer, of course, is also no.

Then why did they sign?

This very question was much of the motivation for this paper, and hence was the first question in a survey of the letter's earliest signatories that we ran over the past three weeks. FLI's open letter acts as a huge inflection point that has helped make risks in AI almost as mainstream as AI itself. It also acts as a record of hundreds of experts who are professionally and emotionally deeply invested in AI and paying close attention to its dangers: each with unique expertise and experiences that led them to sign the letter. We sought to understand signatories' personal perspectives, how their beliefs relate to the letter's stated goals, and to develop a comprehensive understanding of the world's AI experts' deepest sources of distress about their field. We were able to bring clarity to an outstanding range of perspectives behind the letter's initial signatories and to truly unpack the many risks over which the world's most concerned AI experts lose sleep.

## 1.5. Methods

We personally reached out to the first one hundred signatories of the letter with publicly available emails. In hopes of maximizing our response rate, we also emailed nearly every early MIT-associated signatory (since our matching addresses would certainly reach their inbox). We asked each signatory for a 30-minute interview to discuss the letter, their alignment with its contents, their perspective on AI today, and their hopes for AI's trajectory. We also developed two Google forms (see Appendix) for signatories who didn't have time for an interview but still wished to share their thoughts. From 143 emails, we received 37 contributions and had the opportunity to speak personally with 21 signatories. Though suffering from an exceptionally small sample size and unabashed sampling bias, our survey benefits from a diversity of background: interviewees spanned five continents and ranged from retired academics who saw AI develop from perceptrons, to seasoned and active researchers, to professors of law, to industry CEOs.[2]

## 1.6. Findings

Though most of these signatories were among the letter's first, almost none of them signed with considerable hope for a six-month pause of large AI model development. More than a few didn't believe a six-month pause was realistic or even the right step. Nor did a large majority "envision the apocalyptic scenario that some parts of the document warn about" (Anonymous, 2023). Our interviewees all agreed with the letter to some extent, and they all thought it was valuable, but it was evident that these experts' perspectives are as diverse as the larger public's. Our contributors' consensus was summarized by Benjamin Rosman[3]: "Obviously you don't agree with everything when you sign a document like this. You weigh it up. And on average the sentiment is aligned with what I believe" (Rosman & Kupiec, 2023).

Within this paper, we outline the most relevant discoveries from our conversations with signatories, describe their general motivations for signing, and dive deeper into some of the greatest concerns that drove their decision to sign the letter. We describe a range of motivations behind their signatures, the many worries about the state of AI today (rather than just existential fears) that drove those motivations, showcase some of their greatest hopes and excitements about these exciting tools, and share some of their ideas for where to go from here.

## 2. Why Sign?

The most obvious type of signature, implicitly applied to most of the letter's signatories, is one given with complete faith in the validity and realism of a pause in large AI model development. Many interviewees truly identified with the letter's call for a pause. One anonymous respondent "struggle[s] to see any point in history where a six-month cool-down wouldn't have improved tech adoption" (Anonymous & Struckman, 2023). Peter Reiner[4] thinks it would be "a particularly wise decision" to "take a step back and give everybody a chance to breathe a bit" (Reiner & Struckman, 2023). And Simon Mendelsohn[5] feels "slowing down the development of such extremely powerful AI systems will give humanity more time to adequately prepare" (Mendhelson, 2023). But despite sympathies with the pause, no signatories we spoke to had strong confidence in its realism.

---

[2] To be crystal clear, the innate selection bias of all this document's comments is that they are from people who signed the letter; the comments therefore do not include the opinions of those in the AI community who disagreed with the caution the letter proposes or those agreed with its proposed caution but not with its solution and/or reasoning.

[3] Benjamin Rosman is a Professor in the School of Computer Science and Applied Mathematics at the University of the Witwatersrand, South Africa, where he runs the Robotics, Autonomous Intelligence and Learning (RAIL) Laboratory and is the Director of the National E-Science Postgraduate Teaching and Training Platform (NEPTTP).

"I did not sign the letter because I was expecting or hoping that there will be a six-month pause, and quite frankly … I don't believe anybody that signed that letter would actually expect a six-month pause" (Perilli & Struckman, 2023). Alessandro Perilli[6] had it right. Very few interviewees chose to sign the letter because they thought it might lead to the intended pause. And though some identified with a call for a pause, few claimed to sign because they agreed with all or even most of the letter.

- Ricardo Baeza-Yates[7] "thought that the request was not the right one and also that the reasons were the wrong ones" (Baeza-Yates, 2023).
- Moshe Vardi[8] "disagreed with almost every line" (Vardi & Struckman, 2023).
- An anonymous signatory "didn't read it all and [doesn't] buy into it all" (Anonymous & Struckman, 2023).

So, what did convince our interviewees to sign? Sections 2.1 - 2.4 enumerate their major motivations.

## 2.1. We Need to Slow Down

As Microsoft launched its ChatGPT-powered Bing search platform, CEO Satya Nadella announced, "a race starts today... we're going to move, and move fast" (Chow & Perrigo, 2023). Proclaiming "Code Red," a member of Google's management team echoed this sentiment, shifted the company's focus toward AI projects, and instated a fast-track review process for expedited releases (Roose, 2023). Big tech seems to believe it must move fast or be left behind. Rosman is one of many who thought optimizing for speed is the wrong approach to technology that is "qualitatively different to any other problem we've tackled before" (Rosman & Kupiec, 2023).

---

[4] Peter Reiner is Professor in the Department of Psychiatry at the University of British Columbia, a member of the Centre for Artificial Intelligence Decision-making and Action, and founder of the Neuroethics Collective, a virtual think tank of scholars who share an interest in issues of neuroethical import.
[5] Simon Mendelsohn is a Software Developer at the Broad Institute of MIT and Harvard. He previously worked at Amazon.
[6] Alessandro Perilli is a former Fortune 1000 technology executive with 23 years of experience in business and product strategy for emerging tech and open source. Since 2023, he's the CEO of the UK research and development firm Unstable Reality and author of the newsletter Synthetic Work, which is dedicated to the impact of artificial intelligence on jobs and human labor.
[7] Ricardo Baeza-Yates is currently Director of Research at the Institute for Experiential AI of Northeastern University, Silicon Valley campus, since January 2021.
[8] Moshe Y. Vardi is University Professor, Karen Ostrum George Distinguished Service Professor in Computational Engineering at Rice University, where he is leading an Initiative on Technology, Culture, and Society. He is also a Faculty Scholar at the Baker Institute for Public Policy at Rice University.

Several interviewees felt that software developers have spent decades in an industrial environment that encouraged speed of progress over caution. Phillip Isola[9] aptly described the motivation: "It can be the fastest way to solve problems… [and], to be fair, it's very hard to anticipate [consequences] before they arrive. A lot of people have a pragmatic perspective that you need the first catastrophe to regulate it" (Isola & Struckman, 2023). But Rosman feels this perspective is a dangerous one to hold in the face of a new class of software. Typically, "you iterate, and you see when good things and bad things happen, and you revise." But with today's AI, "there is a potential branch of the future where you only have one shot at getting something right. And I think that means the stakes are higher there" (Rosman & Kupiec, 2023). Rosman was not alone.

For varied reasons, every signatory we heard from agreed that the unfettered speed of AI development today "is probably not the right model for the present time" (Isola & Struckman, 2023). Though many signatories agreed a six-month pause was neither realistic nor the best solution, they largely appreciated the letter's understanding of the "importance to go public" with concerns about AI's speed of progress (Shneiderman, 2023). Each signatory's unique perspective on the most critical concerns in AI today (on which we will speak later) drove their prioritization of a public statement. They signed the letter to publicly grab the attention of and encourage caution in three distinct groups: the developers of large AI models, regulators, and the public.

## 2.2. To Developers: Here's an Out

It was evident throughout our interviews that many signatories did not demonize the teams developing today's state-of-the-art models. Some, technologists themselves, related to the developers. Isola thinks "this is all happening because of competition. It's not just money but market and evolutionary forces and incentives toward power" (Isola & Struckman, 2023). By slowing down in this heated environment, companies could put themselves at a major disadvantage. "These companies are not actively interested in destroying the world," Kuipers explained, "but the probability [of a pause] is probably not high partly due to [companies'] lack of trust" that others will do the same (Kuipers & Kupiec, 2023). Though only Max Tegmark mentioned Moloch by name, it is clear that some signatories largely blame the systemic pressures he represents for the current trajectory of AI.

Some signatories hoped the letter could help fight highly influential and Moloch-fueled forces that drive developers of large AI models to push forward relentlessly. Reiner believes "the pause letter gives some cover to companies for being cautious." He adds that "it's impossible to know whether they would've rushed even faster than they're rushing now or whether they've really slowed down. But I think they have" (Reiner & Struckman, 2023).

Not all interviewees had such faith in the letter's influence on developers. Perilli strongly feels "we're talking about the once-in-a-lifetime chance to be the person that delivers Artificial General Intelligence … Do you really think the people that realize that will let a group of researchers or AI ethicists get in the way?" (Perilli & Struckman, 2023). As a result, several signatories had additional motivations upon signing: to get the attention of the regulators who can slow developers.

---

[9] Phillip Isola is an associate professor in EECS at MIT studying computer vision, machine learning, and AI. Previously, he spent a year as a visiting research scientist at OpenAI, and before that he was a postdoctoral scholar with Alyosha Efros in the EECS department at UC Berkeley.

## 2.3. To Regulators: It's Time to Act

Governments and regulators are notoriously snail-like relative to most organizations and can be practically motionless relative to big tech. Many hoped the letter would combat this and introduce necessary, hastened dialogue at a regulatory level. Alan Winfield[10] signed because "we need to draw attention to AI's extraordinary risks. A thousand signatures won't make a difference. Regulators will" (Winfield & Struckman, 2023). Alessandro Saffiotti[11] agreed. "I hoped to raise awareness about the dangers of this uncontrolled development and use of LLMs, both for the general public and the politicians" (Saffiotti, 2023). Perilli felt the letter was "the only way to attract the attention that we have attracted in terms of media coverage, in terms of big questions asked, in terms of regulators paying even more attention" (Perilli & Struckman, 2023).

Not all signatories have high hopes. As John S. Edwards[12] commented, "Governments generally don't understand tech except as a generator of tax revenues (and they can't even make that work... tech companies outsmart them without needing AI)" (Edwards, 2023). Isola, alternatively, is "optimistic about how quickly things are moving in the US government. The Biden administration has been realizing it's very important. They'll probably mess it up, but it's better than ignoring it" (Isola & Struckman, 2023).

Whatever their feelings about either developers or regulators, it seems not a single interviewee signed without some hope for improving (however they define improvement) the mainstream perception and understanding of AI today.

## 2.4. To the Public: Pay Attention

The open letter undoubtedly ignited mainstream dialogue on the world's mass adoption of poorly understood AI tools. "Poorly understood" applies in some key ways to developers who can't interpret their products. But it also applies far more broadly to the general public. Today, public perception of AI is split between two extremes, both of which severely lack nuance:

In the eyes of the public, AI is a vaguely but terrifyingly destructive construct, shaped by media which sensationalizes it as dystopian and anti-human. It is simultaneously a shiny new object to chat with, get homework help from, and automate basic tasks. It will someday produce such economic value that we'll find ourselves in a utopia. Or it will kill us all.

---

[10] Alan Winfield is a (semi-retired) Professor of Robot Ethics at the University of the West of England (UWE), Bristol, UK. He is a Visiting Professor at the University of York and an Associate Fellow of the Centre for the Future of Intelligence, University of Cambridge. Winfield co-founded the Bristol Robotics Laboratory, and his research is focused on understanding the nature and limits of robot intelligence.

[11] Alessandro Saffiotti is a full professor of Computer Science at the University of Örebro, Sweden, where he heads the AASS Cognitive Robotic Systems Laboratory. He has been active for more than 30 years in Artificial Intelligence (AI), especially in the integration of AI and Robotics.

[12] John S. Edwards is an Emeritus Professor in the Operations and Information Management Group at Aston Business School.

These are caricatures of public perceptions of AI, but most interviewees did feel a significant gap between reality and mainstream understanding. Chuck Anderson[13] explained that he has "been concerned for years about the state of AI research promoted by… publications that are not sufficiently reviewed and by the popular press. [It's led] to a misunderstanding of the capabilities of AI and to a lack of consideration of potential harm" (Anderson, 2023).

Interviewees were driven to be a part of something they expected to help resolve this disconnect and make a splash in important ways. Baeza-Yates knew that "given the main signatories, [the letter] had a potential for public impact" (Baeza-Yates, 2023). Vardi recognized the letter as the "first public statement from many tech leaders–including industry–saying AI is too important to leave it to the free market" (Vardi & Struckman, 2023).

Joe Kwon (whose research helps solve immediate problems by improving the ability of models to behave as expected in novel situations) knew that he'd "be happy with side-effects of the letter calling public attention to risks and proposed regulations even if the moratorium failed" (Kwon & Struckman, 2023). And James Koppel[14], who worries most about AI's existential risks, would do a lot to "convince one person to take AI risks more seriously" (Koppel & Struckman, 2023). Kwon and Koppel have wildly different perspectives on AI's most pressing issues and therefore well represent the impressive range of interviewees who sought to capture the public's attention with the letter. As Koppel commented, the letter aimed to "get as many people on board as possible with some common ground" — a goal it certainly achieved (Koppel & Struckman, 2023).

The letter brought each signatory together, but the concerns behind their signatures varied incredibly. FLI's letter describes existential risks in AI that are not definitively shared by the expert signatories we spoke with. "Doomsday is not upon us," Vardi asserted. "[I'm] not saying whether it's coming or not, but now we must deal with the clear and present danger" (Vardi & Struckman, 2023). While a few aligned with the letter's existential focus, many, like Vardi, were far more preoccupied with problems relevant to today. We spoke deeply with each signatory about the immediate and non-theoretical facets of AI that most disturb them.

## 3. Why Slow Down?

Whatever finally motivated their signature, each interviewee was driven by serious concern over the state of AI. Their diverse expertise meant this serious concern was often focused on different aspects of the field. The risks they discussed ranged spectrums between AI today and in the far future as well as between AI's development and its effects. But they all stemmed from the same source.

Stuart Russell, Professor of Computer Science at UC Berkeley, and public signatory of the letter, recently asked senior OpenAI employee, Sébastien Bubeck, "whether GPT-4 has developed its own internal goals and is applying them in choosing its outputs." The answer reflected the root of nearly every interviewee's fears: "We have no idea" (Russell, 2023).

---

[13] Chuck Anderson is a professor in the Department of Computer Science, Colorado State University (CSU), working on machine learning algorithms and deep neural networks for reinforcement learning, brain-computer interfaces, and high-dimensional data in general.

[14] James Koppel's focus is on improving the world's software quality. He founded a business training professional software engineers how to write better code and previously performed research on advanced programming tools at MIT, where he completed his Ph.D. in programming languages.

Sections 3.1 to 3.3 represent our categorization of our interviewees' concerns about AI.

## 3.1. The Root of the Problem

Bubeck's response validates many signatories' perspectives, including Arvind Tiwary's[15] claim that state-of-the-art AI systems today "are engineered with trial and error and no theory" (Tiwary, 2023). Claims like this are driven by rapid and untested deployment of advanced models that are uninterpretable and potentially misaligned.

### 3.1.1. Interpretability & Alignment

Today's state-of-the-art models, including GPT-4, "are so uninterpretable, we have often no idea what they're doing. You could also say experts across fields still have a lot to learn about the way these large language models work" (Rojas & Kupiec, 2023). Camilo Rojas'[16] words were echoed by many interviewees. Developers can only see the inputs and outputs and have little to no understanding of the inner workings of the algorithm. How is it making decisions? Why is it making them? Quite literally no one entirely understands the answers to these essential questions.

But who cares? It works, right?

Signatories like Barry Bentley[17] believe the current lack of interpretability of 'black box' AIs doesn't allow for "the internal reasoning and mechanisms [to] be scrutinized" (Bentley, 2023). Saffiotti chimed in similarly: "We are building systems beyond the level of complexity that we can understand and control. Despite that, we might end up using them in ways that have a direct impact on the life of humans and the planet. This is an extremely dangerous situation" (Saffiotti, 2023).

The dangerous situation Saffiotti refers to arises when a common issue in machine learning, the alignment problem, is scaled globally. Because we don't fully understand why they do what they do, it can be exceptionally difficult to tell complex AI systems what we want from them. If the AI is designed to play a racing game and we tell it to maximize its score, it might amusingly travel in circles, collecting regenerative coins but making no course progress. If we direct a much more powerful system with influence beyond a video game, it's hard to know what it might do.

Andrew Barto[18] talked about this, referencing Norbert Wiener, the father of cybernetics, "in the last section of [his] book on reinforcement learning.… You ask for something, but not what you should have asked for, or what you really want. The system will come up with something that optimizes the objective function that you've specified, but you don't know beforehand what the optima are going to be. And that's a problem" (Barto & Struckman, 2023).

---

[15] Arvind Tiwary is Chair Emeritus TiE Matrix Forum and Founder Sangennovate.
[16] Camilo Rojas is an SNSF Fellow in the Fluid Interfaces group at the Media Lab, working with Pattie Maes on machine learning and wearable devices for the new generation of empathetic technology.
[17] Barry Bentley is a Senior Lecturer in the Cardiff School of Technologies. He completed his Ph.D. at the University of Cambridge, where he worked at the MRC Laboratory of Molecular Biology in the areas of computational neuroscience and bioinformatics.
[18] Andrew Barto is currently Professor Emeritus of computer science at University of Massachusetts Amherst. He is best known for his foundational contributions to the field of modern computational reinforcement learning.

Concerns over system interpretability (which lead to worries over alignment) are not new. Interpretability and alignment difficulties are the driving force behind arguments for AI's existential risk. If we don't know why it does what it does, and we don't know how to tell it exactly what we want, how can we safely give it any power?

The immense number of unknowns involved with complex AI systems today genuinely frightens many of our interviewees. And these fears have led seasoned computer scientists to seriously reconsider their field's cultural celebration of unfettered exploration and innovation. Winfield stated his position on this approach simply: "All of this is just bad engineering" (Winfield & Struckman, 2023).

### 3.1.2. Bad Engineering

Winfield is particularly disturbed by the AI community's willingness to deploy products it doesn't understand for the sake of profit: "Cautious engineering takes time, and we need cautious engineering. 'Move fast and break things' is the worst thing I've ever heard. Imagine if we said that in the railway business" (Winfield & Struckman, 2023).

Barto shared this worry: the "AI community is not quite sensitive to some of the standard engineering practices. If a civil engineer comes up with some new way to build a bridge, they won't just build a bridge and let the public try it. It's not what engineering is about" (Barto & Struckman, 2023).

These interviewees, many experienced developers themselves, believe the technology community's "leap first, look later" culture is entirely inappropriate for AI. Some sympathized with the previous "move fast and break things" approach while others' hindsight determined it was always faulty, full of hubris, and destined to hurt underserved populations. But almost all signatories we heard from, whatever their perspective on the past, thought it was an entirely unsafe culture in which to develop today's powerful AI systems.

The race for supremacy in the AI market is Moloch-driven and incredibly difficult to slow. But between difficulties understanding and controlling powerful AI systems and the questionable engineering practices under which they're deployed, signatories we spoke with were highly troubled about state-of-the-art models' potential to be both intentionally and unintentionally misused.

## 3.2 Direct Effects

Implementing any new technology inevitably introduces unintended negative side effects. But interviewees fear the current trajectory of AI could cause extraordinarily unnecessary and impactful issues. Systems controlled by Moloch encourage developers to do anything for an advantage in the race for AI supremacy, and avoidable harms can be created or exacerbated by powerful, rushed, and poorly understood AI systems. Rosman feels industry's approach to AI deployment "is pretty reckless at this point. Even when they see that their demos and prototypes are doing ridiculous things, they double down on it" (Rosman & Kupiec, 2023). Many signatories are extremely troubled about a myriad of problems these "recklessly" deployed models will cause.

### 3.2.1. Misinformation & Manipulation

In 2019, OpenAI first released GPT-2, an LLM capable of writing short coherent outputs. They held off releasing the full version out of "fear it would be used to spread fake news, spam, and disinformation" (Vincent, 2019). An internal study determined "GPT-2 can be find-tuned for misuse," but there was "no strong evidence of misuse so far" (OpenAI, 2019). Every subsequent LLM has suffered from more of these problems. Previous models pale in comparison to today's state-of-the-art, and multiple signatories believe misinformation proliferation caused by AI is of the field's most relevant and urgent risks.

LLMs excel at producing long, coherent text. Unfortunately, they often generate complete falsehoods that can be extremely convincing to non-experts. Quoting Byrne Hobart[19], author of the technology and finance newsletter "The Diff", one anonymous respondent told us "LLMs are better at sounding informed than being informed" (Hobart, 2023).

In a new study, Georgetown University, OpenAI, and Stanford University identified that LLMs can be leveraged to produce highly persuasive, easily scalable campaigns to "covertly influence public opinion" (Goldstein et al, 2023). Barto worries this influence will be applied by propagandists using AI to further their agendas and "manipulate the political landscape" (Barto & Struckman, 2023). Attempts at political and social manipulation using generative AI are called "influence operations" and already occur at some scale (Bateman et al, 2021). Perilli echoed this apprehension; "Depending on how these models have been trained and fine-tuned, they can expose certain kinds of bias in terms of your political affiliation. When you are exposed to the recommendation generated by an AI that is leaning in a certain direction, you tend to have your behavior or your line of thinking influenced in this way" (Perilli & Struckman, 2023).

Vardi's fear of disunity sparked by AI-driven algorithms is even stronger: "we now see different realities and are experiencing not just cognitive polarization but also economic polarization." Vardi was clear that the issue isn't new, and "AI already runs the world in many ways" (Vardi & Struckman, 2023). Powerful AI systems will not introduce widespread misinformation to the world, but they may certainly exacerbate it. AI-driven algorithms have always shown us "information that will maximize profit," and as they become more complex and less interpretable, it will become increasingly likely that profit-maximizing content does not benefit individuals, society, and (in the long run) even the algorithm's developers (Vardi & Struckman, 2023). AI-generated (rather than just AI-recommended) content could easily add to this problem.

Emotional manipulation by powerful AI systems doesn't have to occur at global scales to be a cause for unease. Several interviewees also cited risks caused by the relative ease with which individual humans grow emotionally attached to chatbots.

---

[19] Not a signatory of the FLI Pause Letter.

### 3.2.2. Unhealthy Attachment

The Eliza Effect refers to the tendency of users to attribute human-level intelligence and emotions to AI chatbots. It was identified in the 1960s when people began to develop intimate relationships with MIT professor Joseph Weizenbaum's simple chatbot, ELIZA (Glover, 2023). According to Perilli, "humans have not changed since those early interactions with ELIZA. Rather, what has changed is the power of AI" (Perilli & Struckman, 2023).

Rojas argues "human-like quality [in AI systems] is unnecessary at this point" and even dangerous. "We are not being mindful of the consequences of [an AI] telling people information as if it was a human" (Rojas & Kupiec, 2023). Rojas described recently developed wrappers around ChatGPT that present the technology as a "coach." To him, this is a dangerous association to make because a coach-coachee bond is extremely strong. Unintended effects of this "coach" could simply influence the coachee's mood (e.g., by unintentionally and unknowingly hurting the coachee's feelings) or even influence the coachee's relationships. Baeza-Yates echoed fears for AI influence and brought up a recent incident where a "person that committed suicide in Belgium" may have been encouraged to do so by a chatbot (Baeza-Yates, 2023).

The Eliza effect was once limited to controlled research environments. Today, as Moloch has us deploy unprecedentedly uninterpretable AI systems at unprecedented rates, several signatories worry it could proliferate wildly.

### 3.2.3. Breakdown of Trust

It's evident that signatories are seriously concerned about many unaddressed side effects of today's state-of-the-art models. Because of this, several are also worried these faulty products might deteriorate societal trust in the systems into which they're integrated. Over time, the United States has shifted from "caveat emptor" ("let the buyer beware"), a belief that it is the responsibility of buyers to ensure a good or service is of their desired quality. Today, robust regulations allow US and many global citizens to believe in "caveat venditor" ("let the seller beware") (Kagan, 2023). They trust that a given product or service will meet certain expectations. Kuipers says "we actually depend on that trust. If we undermine our trust sufficiently, we could do a huge amount of damage to society" (Kuipers & Kupiec, 2023).

This is especially relevant as often-inaccurate LLMs are increasingly used both in content recommendation and generation. Barto explained that humans currently rely on the on-average validity of their sources. "If a respectable journalist puts something forward, you tend to think that that's true. And if it's not true, they'll retract it" (Barto & Struckman, 2023). But he questions how you implement something like that with a chatbot. Similarly, Edwards fears that AI-perpetuated disinformation will get to the point where "no-one can believe anything that is not said face-to-face, leading to potential major societal issues." He questions, "Can I be sure you are genuine? Can you be sure I am genuine (Edwards, 2023)?

These fears are not unjustified. We live in a world where many receive news through algorithmic recommendations and do not think particularly critically about its source. If unfettered AI-shared or AI-generated misinformation continues to spread, Rosman feels "our democracies can't handle that load of abuse" (Rosman & Kupiec, 2023). Interviewees referred to this kind of abuse as an unintentional side effect of today's powerful AI systems. It is perpetuated and ignored due to systemic complexities at Moloch's command. But not all abuse can be attributed to systems.

### 3.2.4. Bad Actors

Yu-Ting Kuo is an MIT Sloan Adjunct professor who founded Microsoft's Azure AI Cognitive Services and Computer Vision Groups and just retired as Microsoft Corporate VP in 2021. Though he did not sign the open letter, he kindly spoke to us about his experiences deploying powerful AI technology over the past two decades. "We used to argue that tech is neutral: there are bad actors and good actors. That was my excuse." He described how excited his team was by technological democracy when Microsoft published one of the first first open-source facial recognition systems in 2015. "We did not anticipate or think of the technology falling into the hands of government agencies using it to harm citizens. I regret it. We didn't think of it" (Kuo & Struckman, 2023).

Today, with a tiny investment, anyone can access GPT-4's API and use one of the world's most powerful AI systems for any purpose. Isola, a lifetime proponent of open-source technology, told us "I find it deeply uncomfortable to think about whether or not these should be open-sourced" (Isola & Struckman, 2023).

In particular, Vardi fears that phishing emails are about to escalate in both quantity and quality. "How many people know how to look for phishing today? As it gets better faster, how quickly will that percentage drop?" He feels, given open-source access to powerful APIs like GPT-4, "that horse has already left the stable" (Vardi & Struckman, 2023). Rosman worries another well-known fraud, false kidnapping claims for ransom, will explode in effectiveness. Using modern generative AI, anyone with access to a short snippet of your voice can clone it for any purpose (Rosman & Kupiec, 2023). Rosman fears the many ways "this [tech] can be maliciously used, which could be anything from scamming people out of huge amounts of money to potentially producing toxic agents that could kill large amounts of people" (Rosman & Kupiec 2023). He cited a recent study where researchers used LLMs to synthesize Ibuprofen–a cool discovery that could be nefariously applied to more dangerous substances with little effort (Boiko et al 2023).

Isola is "most worried about the weaponization of this technology. It seems like a sure thing that a lot of countries will be tuning and applying these models to war. Whatever country has the most powerful AI has an advantage, and that creates an arms race" (Isola & Struckman, 2023).

Many interviewees were unhappily confident that powerful AI systems hold prolific and disturbing risks for individuals and societies alike. Rosman feels we are doing a terrible thing by "just opening the doors to make it easier to produce ways of harming other people" (Rosman & Kupiec 2023). Millions in vulnerable populations are at the mercy of anyone with bad intentions, some limited capital, and a little technical ability. Bad actors must meet an exceptionally low bar to abuse the most powerful technology in history. This bar is only so low because the incredible speed at which Moloch forces companies to deploy new powerful AI models inhibits their ability to carefully mitigate misuse. This mismatch in speed makes it both needlessly and dangerously simple for "speculators…waiting to exploit whatever comes to their hands with no legal nor ethical concern" (Anonymous, 2023).

## 3.3 Indirect Effects

Incredible technological advancements tend to cause ripples of profound cultural and economic changes. The first industrial revolution allowed humans to mechanize production using water and steam. We soon developed mass production and harnessed electricity. The recent digital revolution led to electronics, the internet, and automation. Each of these technological advancements quite literally revolutionized the world, and the new tools had effects far beyond their immediate scope.

Now we are on the crux of the "Fourth Industrial Revolution", a term coined in 2016 by Klaus Schwab[20], founder of the World Economic Forum (Schwab, 2016). Powerful AI systems of this new revolution could have significantly greater global impact than any previous technology, and at a much more rapid pace.

Interviewees felt many of these changes can and will be exceptionally positive. But several were preoccupied by the speed and unwieldiness of this fourth revolution. Technological revolutions all share similar downsides. Jobs are displaced, underprivileged populations are taken advantage of, value is poorly distributed, and the environment suffers. What happens when the pace and power of changes are orders of magnitude greater than even the digital revolution? Some interviewees fear the scale of change will be greater than society is ready for. Job displacement, the most mainstream of these four effects, is at the top of many minds.

### 3.3.1 Job Displacement

In a recent report, Goldman Sachs estimated that 300 million jobs could be lost or diminished by the proliferation of AI (Kelly, 2023). Samuel Tenka[21] emphasized that today's state-of-the-art models are already capable of extraordinary displacement. "Much of what the creative class does, surprisingly, can be automated. [Today's AI] is not anywhere near human cognition. But, in many jobs it just needs to be good enough. This hits closer to home because I write and do math… I feel vaguely threatened" (Tenka & Kupiec, 2023).

It is globally evident that modern AI will severely alter the job market. "It's going to be a larger transformation than probably has happened to humanity in all of history, in terms of jobs. Not necessarily just that they'll all disappear, although some will, but all sorts of new niches will open up" (Reiner & Struckman, 2023).

But will they open fast enough? Rosman, in particular, thinks it is unrealistic to expect society to carefully upskill the newly unemployed into emerging roles (Rosman & Kupiec 2023). Perilli agrees. "I've never felt, as a technologist with deep technical expertise, any risk in terms of serious impact on the jobs… Here, not only do I feel this risk, but it feels so tangible and so real… [This] comes from a person that, until four months ago, represented one of the biggest automation projects in the world (Ansible at RedHat)" (Perilli & Struckman, 2023). Vardi emphasized that it is easy to ignore the portion of the working class that's been left jobless by automation, and he's extraordinarily concerned that today's powerful systems are ready to tackle more creative tasks and infiltrate white collar jobs (Vardi & Struckman, 2023).

---

[20] Not a signatory of the FLI Pause Letter
[21] Samuel Tenka is a Ph.D. student at MIT's Center for Brains, Minds & Machines.

Historian and letter signatory Yuval Noah Harari introduced the idea of "the useless class" years ago. Mass industrialization created the working class, and Harari hypothesizes the AI revolution will create a new unworking class that "will not merely be unemployed — it will be unemployable" (Harari, 2017). Alejandro Bernardin[22], generally optimistic about the trajectory of the job market, describes how "you can think [about the useless class] in two ways. It's very good because we can do human things, for instance, to play, to sing, to dance, to be with the family, right? And the other way: I will have no job. Who will pay me? Where am I going to get my money to eat" (Bernardin & Kupiec, 2023)?

To Rosman, this could be exceptionally dangerous: "The problem is, with a huge unemployment rate, a chunk of your society is not invested in the success of the society… This could be destabilizing to large economies. And destabilizing in a way that could end with animosity and potentially violence" (Rosman & Kupiec, 2023).

Beyond destabilization, signatories are worried about the effects of "cognitive offloading." Reiner used this term to describe how these tools will allow and encourage humans to offload increasingly complex tasks like drafting summaries, writing code, and more. Cognitive offloading is not inherently bad (e.g., an online calendar is much easier than mentally keeping your schedule), but if state-of-the-art AI systems continue to improve, Reiner feels it's unclear what may remain for the average human (Reiner & Struckman, 2023). With all this extra time, Harai questioned "what will keep [humans] occupied and content? One answer might be drugs and computer games… What's so sacred about useless bums who pass their days devouring artificial experiences" (Harari, 2017)? Another anonymous signatory who works in rehabilitation, worries about "'brain-disuse syndrome' if generative AI is not well used" (Anonymous, 2023). Reiner wonders if society might be moving towards a world like that in the movie *Wall-E,* and if we'll respond differently than the happily inert human characters it depicted (Reiner & Struckman, 2023).

Uneasiness over job displacement was mentioned by most interviewees to varying degrees of severity. There was no consensus over how incoming changes in the job market will affect society, but all respondents were confident in and unnerved by these changes' extremity and rate of approach. Whatever our approach to powerful AI systems going forward, job displacement is already a major inevitable effect of today's state-of-the-art AI.

Ironically, certain jobs that AI development *does* create are key pieces in a different concern some signatories hold about the AI revolution.

### 3.3.2. AI Colonialism

AI colonialism is a term increasingly used by scholars who believe "the impact of AI is repeating the patterns of colonial history" through cheap labor exploitation and, less directly, cultural imperialism (Hao, 2022). Multiple interviewees told us they were agitated about both the unethical ways AI development abuses underprivileged employees and the cultural influence powerful AI systems are destined to have.

---

[22] Alejandro Bernardin is a postdoctoral researcher at the Computational Biology Lab (DLab), a transdisciplinary laboratory for fundamental and applied research belonging to Fundación Ciencia y Vida (FCV), Santiago, Chile.

Winfield was particularly angry at the egregious state of outsourced content moderation jobs. Content moderation algorithms are key to reviewing millions of daily posts on online platforms. But since these algorithms struggle with nuance, they can be taken advantage of by clever bad actors changing a few letters or pixels in offensive text or images. The meaning is still clear to human readers, but it flies under the radar of algorithmic moderation. Supplementary human moderation is therefore still essential to online platforms, and these jobs are often outsourced to third world countries with relatively low standards for employee wellbeing. Days before FLI's letter was published, Kenyan content moderators sued Meta for unlawful dismissal after unionization attempts (Espada, 2023). Winfield was particularly disturbed that these attempted unionizers are not only paid a minuscule fraction of the average Meta employee, but also given no psychological support for the traumatizing and disturbing content they engage with for hours a day as a part of their job (Winfield & Struckman, 2023). Interviewees also voiced concern for employee treatment in all aspects of data labeling for AI.

Labeling the data used for training massive AI systems is extremely intensive manual labor that is consistently outsourced to countries where lower wages and limited employee support is accepted. Kuo tells us OpenAI pays its labelers $2/hr and that he wouldn't be surprised if it also does not offer them psychological support (Kuo & Struckman, 2023) (Perrigo, 2023). These human rights issues are a straightforward abuse of power that is reminiscent of historical colonialism, but some signatories mentioned a more subtle concern caused by the global influence of powerful models like GPT-4.

"I don't want to end up with the entire world consuming an AI inside Microsoft Office that is being shaped after, no offense, a 20-year-old white and entitled Caucasian guy that works in the top university" (Perilli & Struckman, 2023). Perilli is worried about the individually and culturally influential effects new models might have because the teams developing them have inherently limited scope. "I just don't want me or you or whoever to be the one that decides how the model has to respond to people all around the world" (Perilli & Struckman, 2023).

When speaking on his 25 years helping lead Microsoft, Kuo reflects that "we didn't realize, and many don't even today, how our values are directly imbued into the product" (Kuo & Struckman, 2023). Inevitably, development teams shape products in their own image; all products suffer from the inherently limited scope of their developers. But what does that mean for powerful AI systems that directly interact with and, as discussed in previous sections, influence the feelings and beliefs of their users? This is why Kuo claims "AI is not a technological problem, but an ideological problem… technology is not neutral" (Kuo & Struckman, 2023) (Lanier, 2020).

As LLMs designed by culturally similar computer scientists are globally circulated, several interviewees are disquieted by the lack of clarity behind how AI systems will interact with and influence different types of people in different ways. But unintentional cultural influence, though adjacent, was not the most discussed issue related to the very small groups that develop today's state-of-the-art models.

### 3.3.3. Power Concentration

As the technological revolution matures, it's become evident that "new technologies have tended to produce winners-take-most outcomes. Dominant firms have acquired more market power, market structures have become less competitive, and business dynamism has declined" (Qureshi, 2022). If AI falls into a winner-take-most scenario, many signatories are troubled by the global influence this could give the winner (or winners).

Reiner frankly offered his stance: "the power on the planet is being concentrated in very few companies that are really on the verge of (they're not quite there) being more powerful than the largest nation states. They're certainly more powerful than the majority of smaller nation states on the planet. We're just walking blindly into that situation. I don't think the companies are unaware of that. The companies are aware of the power that they have now, and that which is coming to them" (Reiner & Struckman, 2023).

Because of this, signatories think the economic benefits of AI will fail to be distributed in any meaningful way. An anonymous signatory told us that "no one's seen the benefit. Trickle-down effects of AI haven't reached the poor people of this country [the UK] any more than it's reached those of yours" (Anonymous & Struckman, 2023). Winfield agrees that the "wealth being created is not being distributed" and isn't pleased that scientific achievements paid for by public money (in the form of endowments and grants) are "now exploited by mega companies making vast amounts of money" (Winfield & Struckman, 2023).

Concentrated power and money are both the result of and the producers of more Moloch-driven forces. Bernardin summed up the worry that many signatories share. "If we continue with this logic, a big company takes all the power and all the jobs, and all the work and all the money. And we don't have a better way to distribute this money" (Bernardin & Kupiec, 2023).

### 3.3.4. Environmental Effects

Interestingly, though training LLMs takes a not insignificant amount of energy, very few signatories addressed concerns they had over AI's effect on the environment. Winfield, however, pointed out the absurdity of pursuing the development of increasingly complex AI amid a climate emergency (Winfield & Struckman, 2023).

Winfield added that though today's AI systems are remarkably capable, they are terribly inefficient. "It cost AlphaGo 50KW to play 2 hours of Go. A human needs an egg sandwich (which is about 5W)" (Winfield & Struckman, 2023). Tiwary directed us to a paper from UC Riverside and UT Arlington which found that "ChatGPT needs to 'drink' [the equivalent of] a 500 ml bottle of water for a simple conversation of roughly 20–50 questions and answers" (Tiwary, 2023) (Li, P. et al, 2023). This is an enormous number considering ChatGPT's billions of users. Environmental concerns are the last and least mentioned in a thorough list of pressing issues these experts discussed. But they, along with every other risk, need to be considered to ensure the unprecedentedly vast and rapid effects of the AI revolution genuinely benefit society.

# 4. What's Going Well?

Though we sought to understand why these experts were worried enough to sign FLI's letter, it would be inaccurate to exclusively present their fears about AI. Every respondent to our survey was excited and hopeful about some aspects of the field (they work in it, after all) and believed it has the potential to be incredibly positive.

Specifically, AI could help solve some critical global challenges. Steve Petersen[23] envisions a future where AI paves the way for groundbreaking achievements such as fusion energy, cancer cures, income redistribution, and resolution of political issues (Petersen & Kupiec, 2023). Isola is passionately excited about AI's applications in biology and drug discovery (Isola & Struckman, 2023). Rojas believes AI has "huge potential for reducing inequality and for making education more accessible around the world" (Rojas & Kupiec, 2023).

Some signatories are fascinated by the nature of intelligence itself and see AI as a route to understanding it. Isola says his lifelong "motivation has been to understand intelligence scientifically. I want to understand humans" (Isola & Struckman, 2023). Reiner is excited that "advances in AI are forcing us as a society to think about what intelligence means…I think it's giving us a lot of opportunity to pause and reflect on our own role in the world, where we sit at the top of the apex cogitate, and what it means when we're no longer at the top of that pyramid" (Reiner & Struckman, 2023).

Simply, signatories believe AI is poised to help us understand the world at large. "I believe in scientific advancement, and I believe and define AI not as a set of applications but as the science of using computational concepts to understand the nature of mind…Understanding the world is better than not understanding the world," Kuipers proclaimed (Kuipers & Kupiec, 2023).

Our interviewees hold multifaceted excitement for AI. It is a tool for addressing pressing global challenges, promoting equality, stimulating introspection, enabling creative expression, and furthering scientific knowledge. The potential for AI to shape a better future and redefine our relationship with the world elicits a sense of wonder and anticipation among these signatories. Whatever their worries, most of them are technologists and enthusiasts first.

But several signatories believe truly positive outcomes from AI applications are only possible if profound changes are made.

# 5. What Now?

After learning the "why" behind these experts' signatures was not unerring agreement with the letter's proposed solution, we asked what they did want for their field. Unsurprisingly, their vision for the next steps toward AI's golden future were as diverse as their worries for it. There was limited consensus due to both varying faiths in capitalistic systems, academic influence, and regulatory bodies, and also differing beliefs in the technical trajectory of future models. However, all their suggestions were focused on necessary regulatory and cultural changes.

---

[23] Steve Petersen is an associate professor of philosophy at Niagara University. His research centers on a strongly naturalistic approach to good thinking and the ethics of artificial intelligence, especially (lately) the problem of the risk posed by superintelligence and value alignment.

## 5.1. Regulation

No signatory thought AI could be fully entrusted to the free market. Winfield wants to quiet myths that "regulation stifles innovation. The regulation of aviation has made it innovative and safe. Regulation doesn't stifle but promotes innovation because it offers frameworks, standards, an agreed way of doing things. It's what allows us to have the internet or to plug our laptop in anywhere in the world" (Winfield & Struckman, 2023). Vardi agrees that "regulation does not stifle innovation necessarily." If it's focused on ensuring safety "it induces trust and helps the market" (Vardi & Struckman, 2023). Regulation is the route to mitigating a breakdown in public trust of AI-driven products.

But interviewees varied in the severity of regulation they felt was needed to face other risks in AI. Perilli heavily prioritized finding a way to encourage economic growth while improving explainability. He made it clear he doesn't "want OpenAI (or any other company) to stop doing what it's doing." It's essential to avoid "the dangerous situation where commercializing an AI model is such a trouble that no one will ever do that." But regulation must create the "capability to scrutinize the dataset and the mechanism to train and so forth" of newly deployed models (Perilli & Struckman, 2023).

Many felt that, at a minimum, limits to public accessibility are needed. Rosman thinks "These ridiculously large models that people are just playing with" don't need to be publicly accessible. "There should be transparency, there should be oversight committees, but I don't think it should be all publicly accessible" (Rosman & Kupiec, 2023). By minimizing public accessibility to models, Rosman hopes bad actors would be slowed in their adoption of this new powerful technology.

Rosman also adds that "it may still be possible to say no model bigger than 'X' should be trained on these [certain domains]" (Rosman & Kupiec, 2023). This might mitigate risks in maintaining model alignment and prevent certain environmental effects of large AI models.

More strongly, Isola claims he "would be more comfortable if [AI development] were just happening in a research context," but clarifies that "it's difficult to separate these things" (Isola & Struckman, 2023). Barto, like Isola, feels strong regulation in industry is necessary but very clearly states "it should not inhibit research. That's kind of a quandary with regulation, that you really don't want to prevent research of these systems" (Barto & Struckman, 2023). Limiting advanced AI development to research purposes would extinguish the Moloch-driven industry rat race for AI supremacy and slow all aspects of the AI revolution: both the good and the bad.

Even further, Vardi, Isola, and an anonymous respondent (with varying degrees of certainty) all mentioned placing advanced AI systems under the same tightly regulated umbrella as nuclear and bioweapons. This would severely slow both industrial and academic development of advanced models, giving society adequate time to study and understand their behaviors. Koppel, who is extremely worried that AI will bring existential risks, does not think AI systems more powerful than GPT-4 are safe to develop at all. He argues that if powerful AI has even a 1% chance of eradicating humankind, we should not pursue it (Koppel & Struckman, 2023).

The diversity of opinions on regulatory severity among these respondents mirrors a global lack of consensus. It seems that the only point upon which signatories are more aligned than the rest of the world is simply that some significant regulation is needed in AI. Our interviewees were far more in agreement over the cultural changes needed in the field.

## 5.2. Cultural Changes

According to a recent TIME magazine piece, "only around 80 to 120 researchers in the world are working full-time on AI alignment" while thousands of engineers are developing more capable AI (Chow & Perrigo, 2023). This proportion is entirely misaligned with reinforcement learning forefather Barto's belief that "explainable AI is a really important thing to be pursued. I must say that in the early days of my work with reinforcement learning it was pretty exciting when the system did things we didn't expect... it doesn't seem so cool when it's available to the world." He feels computer scientists need to adapt to a world that has changed since early days when "we did things in the lab, and it wasn't a product" (Barto & Struckman, 2023).

Barto is one of many well-established academic and industry members who believe AI interpretability and alignment are key topics in computer science. Nevertheless, far more computer scientists and engineers are instead flocking to extending AI capabilities (Chow & Perrigo, 2023). Why?

Anderson is "nervous about how we are teaching the newcomers to the AI field. Many courses just teach the application of deep learning packages, without, what I most value, the details of how they are implemented" (Anderson, 2023). Petersen thinks the tendency of computer scientists to focus exclusively on developing their exciting new technology is becoming increasingly dangerous. "It's the Peter Parker phenomenon. With great power comes great responsibility. But when you work really hard at leveraging your power, you don't stop to think as much about what you want to do with that power." As a result, he thinks "it's worthwhile for schools, even technically focused ones like MIT, to require—I might say some philosophy classes" (Petersen & Kupiec, 2023).

Several interviewees felt that, in response to the speed and magnitude of impacts powerful AI systems will have on the world, a deeper cultural change is needed within the field of computer science. Kuo agreed wholeheartedly as he reflected on his time at Microsoft. Though asking for forgiveness rather than permission can be a "great [mentality] if exercised appropriately," he feels the tech industry is unhealthily used to making all manner of serious mistakes and simply being forgiven. (Kuo & Struckman, 2023).

Our interviewee's proposed cultural changes, much like their regulatory ideas, are far less defined than the issues driving said proposals. Despite their reservations, and some nonsignatories' downright distaste, the FLI open letter must be commended for developing a concrete proposal. Its tangibility successfully brought together some of AI's strongest minds and kindled hundreds of conversations like these.

# 6. Conclusion

As we listened to, read, and parsed the ideas kindly shared with us by these worried experts, we were struck by their simultaneous diversity and alignment. Each interviewee felt differently about FLI's letter, about both existential and contemporary risks in AI, and about viable solutions. But whatever their perspective on the details, each expert we spoke with shared a wish for the field they work in and love to do something unheard of in computer science culture: slow down, forget the tech for a moment, and consider its context. One anonymous respondent's statement captured a sentiment we believe almost all respondents would support:

> "I would not eliminate the benefits of the industrial revolution, but it would have been beneficial to better understand and address corresponding environmental and economic impacts over the succeeding 100 years. I would not eliminate the benefits of the Internet, but it would have been beneficial to better understand and address corresponding impacts on democracy, mental health, privacy, etc. over the succeeding 20 years.
>
> AI holds both promise and risk consistent with industrialization and the internet and, like the internet compared to industrialization, AI is on an accelerated pace.
> Should we not consider the implications more? Should we continue to outsource decisions on how this is deployed to a limited number of tech executives?
> History is to learn from, not repeat" (Anonymous, 2023).

As fledgling computer scientists ourselves, we left these conversations wishing for our field a change in perspective. Technology is inherently incredible. AI is inherently incredible. We do it and ourselves a disservice by giving it up to Moloch, applying it haphazardly, and reaping hollow rewards. Slow down, forget the tech for a moment, and consider its context.

# 7. Acknowledgements

We would not have been able to write this paper without the interviewees who generously gave us the time to learn about their thoughts and careers. We would like to thank (in no specific order) Benjamin Kuipers, Alessandro Perilli, Arturo Giraldez, James Koppel, Andrew Barto, Peter Reiner, Steve Petersen, Samuel Tenka, Joseph Kwon, Camilo Rojas, Alan Winfield, Benjamin Rosman, Christopher Markou, Moshe Vardi, Phillip Isola, Alejandro Bernardin, Yu-Ting Kuo, Simon Mendelsohn, Ricardo Baeza-Yates, Barry Bentley, John Edwards, Ron Kuper, Ben Shneiderman, Alessandro Saffiotti, Chuck Anderson, Arvind Tiwary, Rahi Patel, and those who wished to remain anonymous for contributing their thoughts. Lastly, we would like to thank Alfred Spector for his guidance within this class and sharing his valuable insights.

# 8. Appendix

## 8.1 Short Form Questions

1. What is your technical/academic background?
2. I have a technical understanding of AI. (Rate 1-5)
3. When/how did you first hear about the open letter?
4. What made you decide to sign the open letter?
5. What were your desired outcomes of the open letter?
6. What are your proposed next steps following this letter?
7. I expected the amount of attention the Open Letter publicly garnered. (Rate 1-5)
8. I am satisfied with the public reception of the open letter. (Rate 1-5)
9. I believe that the Open Letter reflected my thoughts as an individual. (Rate 1-5)
10. I believe that signatories of the Open Letter are largely unified on their current perspective of AI. (Rate 1-5)
11. I believe that an AI Moratorium is necessary. (Rate 1-5)
12. I believe that an AI Moratorium will happen. (Rate 1-5)
13. Do you have any additional thoughts on the Open Letter or an AI Moratorium?
14. What excites you most about AI today? About its community?
15. What makes you nervous about AI?
16. I believe that AGI is likely to be achieved within a lifetime. (Rate 1-5)
17. AI is likely to have more positive than negative effects on mankind. (Rate 1-5)
18. Unintentionally misaligned AI is a greater concern than maliciously applied AI. (Rate 1-5)
19. Do you have any additional thoughts on the current state and/or trajectory of AI?
20. Other Thoughts?

## 8.2 Very Short Form Questions

1. I have a technical understanding of AI.
2. I expected the amount of attention the Open Letter publicly garnered.
3. I am satisfied with the public reception of the open letter.
4. I believe that the Open Letter reflected my thoughts as an individual.
5. I believe that signatories of the Open Letter are largely unified on their current perspective of AI.
6. I believe that an AI Moratorium is necessary.
7. I believe that an AI Moratorium will happen.
8. I believe that AGI is likely to be achieved within a lifetime.
9. AI is likely to have more positive than negative effects on mankind.
10. Unintentionally misaligned AI is a greater concern than maliciously applied AI.
11. Other thoughts on any of the following: What made you decide to sign the open letter? What was your desired outcome of the open letter? What excites you most about AI today? About its community? What makes you nervous about AI? Any other thoughts on the Open Letter, an AI Moratorium, or AI in general?

# 8.3 Quantifiable Form Results

(1- Strongly Disagree, 5- Strongly Agree)

I have a technical understanding of AI.

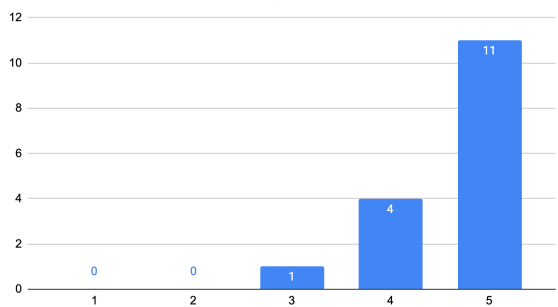

I expected the amount of attention the Open Letter publicly garnered.

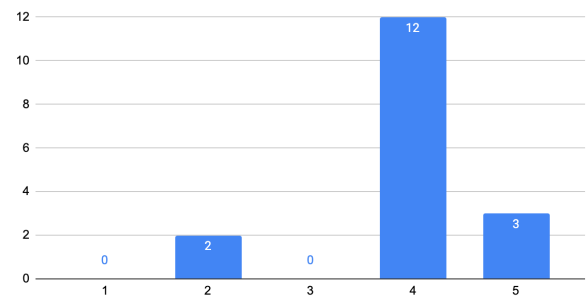

I am satisfied with the public reception of the open letter.

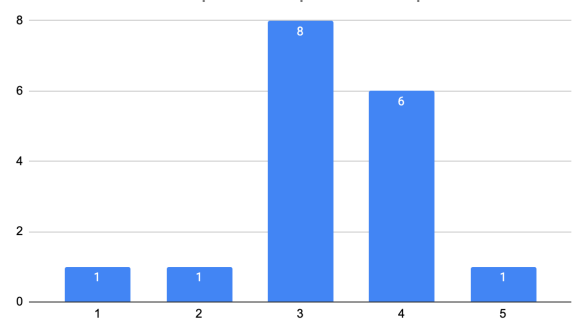

I believe that the Open Letter reflected my thoughts as an individual.

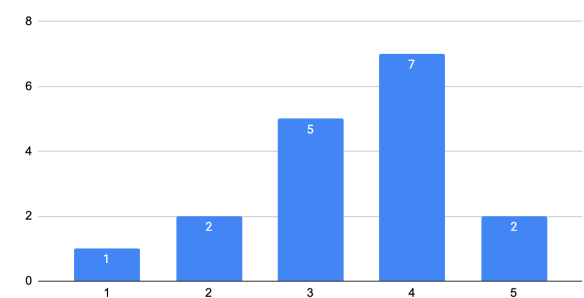

I believe that signatories of the Open Letter are largely unified on their current perspective of AI.

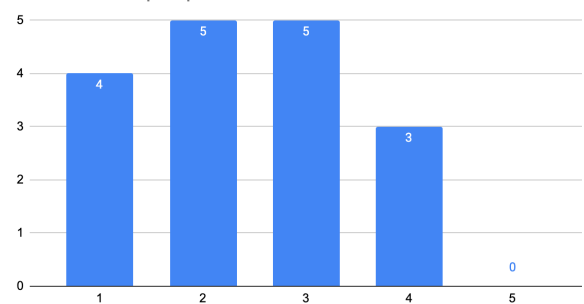

I believe that an AI Moratorium is necessary.

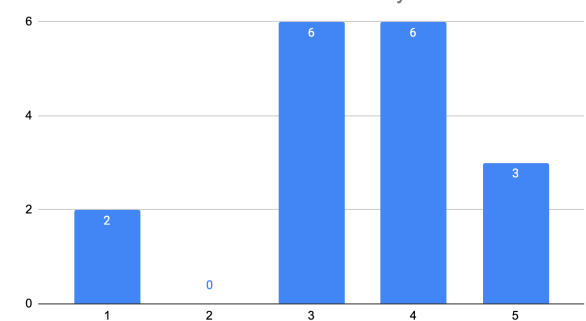

I believe that an AI Moratorium will happen.

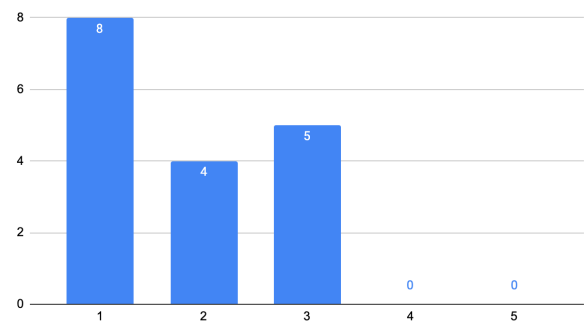

I believe that AGI is likely to be achieved within a lifetime.

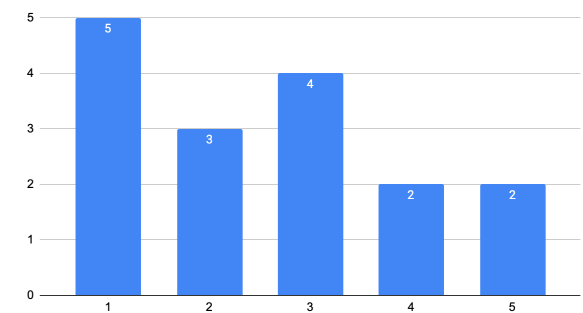

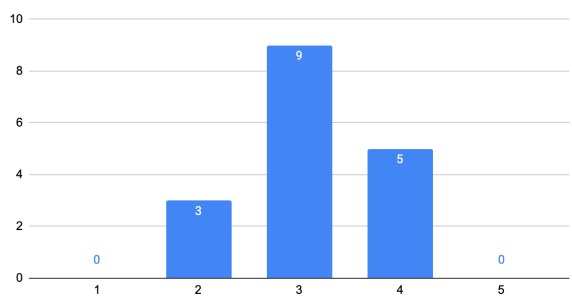

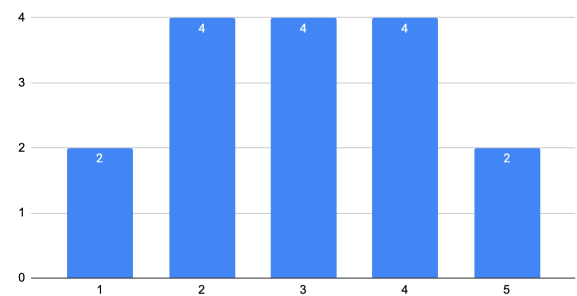